\documentclass[a4paper]{jpconf}

\usepackage{graphicx}

\usepackage{bm}
\usepackage{hyperref}

\usepackage{pstricks,pst-node,pst-text,pst-3d}
\usepackage{amsmath}
\usepackage{amssymb}
\usepackage{cite}

\usepackage{graphicx}

\usepackage{times}
\usepackage[latin1]{inputenc}
\usepackage{textcomp}
\usepackage{color}
\usepackage{rotating}
\usepackage{multirow}
\usepackage{booktabs}

\newcommand{\be}{\begin{equation}}
\newcommand{\ee}{\end{equation}}
\newcommand{\bea}{\begin{eqnarray}}
\newcommand{\eea}{\end{eqnarray}}
\newcommand{\bit}{\begin{itemize}}
\newcommand{\eit}{\end{itemize}}
\newcommand{\bmp}[1]{\begin{minipage}{#1cm}}
\newcommand{\emp}{\end{minipage}}
\newcommand{\bra}{\langle}
\newcommand{\ket}{\rangle}
\newcommand{\bean}{\begin{eqnarray*}}
\newcommand{\eean}{\end{eqnarray*}}

\newcommand{\bfr}{\begin{flushright}}
\newcommand{\efr}{\end{flushright}}

\begin{document}

\title{Lattice QCD at nonzero temperature and density
\footnote{Invited talk at the XXXII IUPAP Conference on Computational Physics 21, Coventry, UK, 1-5 August 2021}
}

\author{G.~Aarts$^{1,2}$, C.~Allton$^1$, S.~Hands$^3$, B.~J{\"a}ger$^4$, S.~Kim$^5$, M.~P.~Lombardo$^6$, A.~A.~Nikolaev$^1$, S.~M.~Ryan$^7$, J.-I.~Skullerud$^{7,8}$
}

\address{$^1$ Department of Physics, Swansea University,  SA2 8PP, Swansea, United Kingdom}
\address{$^2$ European Centre for Theoretical Studies in Nuclear Physics and Related Areas (ECT*) \& Fondazione Bruno Kessler
Strada delle Tabarelle 286, 38123 Villazzano (TN), Italy}
\address{$3$ Department of Mathematical Sciences, University of Liverpool, Liverpool L69 3BX, United Kingdom}
\address{$^4$ CP$^3$-Origins \& Danish IAS, Department of Mathematics and Computer Science, University of Southern Denmark, 5230 Odense M, Denmark}
\address{$^5$ Department of Physics, Sejong University, Seoul 143-747, Korea}
\address{$^6$ INFN, Sezione di Firenze, 50019 Sesto Fiorentino (FI), Italy}
\address{$^7$ School of Mathematics and Hamilton Mathematics Institute, Trinity College, Dublin 2, Ireland}
\address{$^8$ Department of Theoretical Physics, National University of Ireland Maynooth, Maynooth, County Kildare, Ireland}

\ead{g.aarts@swansea.ac.uk}

\begin{abstract}
We discuss some selected recent developments in the field of lattice QCD at nonzero temperature and density, describing in particular the transition from the hadronic gas to the quark-gluon plasma, as seen in simulations using Wilson fermions.
\end{abstract}


\section{Introduction}

The study of strongly-interacting matter under extreme conditions remains a topic of interest. Important questions, e.g.\ present on the list of STFC's Science Challenges \cite{stfc}, include:
 \begin{enumerate}
 \item What are the basic constituents of matter and how do they interact?
\item How do quarks and gluons form hadrons?
\item What is the nature of nuclear matter?
\item Are there new phases of strongly interacting matter?
\end{enumerate}
These questions have been studied over many years, both theoretically and experimentally, since, even though the underlying theory is well known --  quantum chromodynamics (QCD) --, they are nonperturbative and difficult to answer.

Concerning the QCD phase diagram the thermal transition at the physical point is by now well understood \cite{Aoki:2006we,Borsanyi:2010bp,Borsanyi:2010cj,HotQCD:2014kol,HotQCD:2018pds}. The extension to small baryon density is accessible via an expansion in baryon chemical potential over temperature ($\mu_B/T$) and/or via analytical continuation from imaginary chemical potential \cite{deForcrand:2009zkb}. Properties at larger density are harder to access, due to the sign problem \cite{Aarts:2015tyj}, but heavy-ion collisions \cite{Ratti:2018ksb} and compact stars \cite{Annala:2017llu} provide persistent experimental and observational motivation to pursue this direction. In this contribution, the focus is on the thermal transition, and in particular the quark mass (or pion mass) dependence as observed from simulations using $N_f=2+1$ flavours  of Wilson quarks. In addition the onset of parity doubling for baryons is studied, focussing again on pion mass dependence. After presenting some generalities, this contribution mostly follows Ref.\ \cite{Aarts:2020vyb}.
For recent overviews of chiral properties with varying number of flavours, see e.g.\ Refs.\ \cite{Philipsen:2021qji,Kotov:2021hri}.

\section{QCD thermal transition} 

The study of phase transitions in statistical mechanics often follows this standard route: identify an order parameter, discuss which symmetry is broken, and compute derivatives of the free energy (e.g.\ susceptibilities) to see whether and how they diverge in the infinite-volume limit. 
In QCD there are potentially various global symmetries to consider. For massless quarks it is chiral symmetry,
\be
\psi \to e^{i\alpha \gamma_5}\psi,
\qquad\qquad
\bar\psi \to \bar\psi e^{i\alpha \gamma_5},
\ee
 with the chiral condensate $\bra\bar\psi\psi\ket$ as order parameter. However, this symmetry is explicitly broken by the mass term $m\bar\psi\psi$ in the QCD lagrangian.
 
 On the other hand, for infinitely heavy quarks centre symmetry holds, as a symmetry of the nonabelian SU(3) gauge theory. Here the gauge action and path integral measure are invariant under the centre of the SU(3) gauge group, $\mathbb{Z}_3$, when timelike links are multiplied by a phase factor.
The order parameter  in this case is the Polyakov loop, the gauge invariant trace of the product of timelike links. This symmetry is explicitly broken by the presence of not-infinitely-heavy quarks. We conclude that in nature both chiral and centre symmetry are broken explicitly and hence there is no (obvious) order parameter. The transition can then be smooth crossover, with no singularities in the free energy. 
  
Indeed, for physical quark masses it is well established that the transition is a crossover \cite{Aoki:2006we}. Lattice QCD simulations have been carried out  \cite{Borsanyi:2010bp,Borsanyi:2010cj,HotQCD:2014kol,HotQCD:2018pds} directly at the physical point and the continuum limit is taken using at least four values of the lattice spacing, employing the relation
\be
T=1/(a N_\tau): \qquad   a\to 0  \quad \Leftrightarrow \quad N_\tau \to \infty.
 \ee
In practice the continuum limit is approached using $N_\tau = 6,8,10,12,16$.

The studies cited above are obtained using simulations with so-called staggered quarks. These are fast to simulate, have a remnant of chiral symmetry, such that no additive fermion mass renormalisation is needed, but require ``rooting'' to get correct number of flavours (a single staggered flavour realises four tastes, which have to be reduced to one). The correct fermion content is expected to emerge in the continuum limit.

One may ask what the situation is with Wilson-like formulations, if nothing else as an independent check on the staggered results. Conceptually, Wilson fermions do not need rooting,  which eliminates lingering theoretical uncertainties. Since chiral symmetry is broken explicitly, a tuning of the bare quark mass is needed. Wilson fermions are four-component spinors, as in the continuum, and hence more expensive to simulate than one-component staggered quarks.
Hence to date no results in the continuum limit at the physical point have been obtained. Nevertheless, an  emerging consistency is observed, as we discuss here.

\section{Wilson fermions on (an)isotropic lattices}

Studies with $N_f=2+1$ flavours of Wilson quark are available on both isotropic and anisotropic lattices. We denote the lattice size with $N_\tau\times N_s^3$ and the lattice spacings as $a_\tau$ and $a_s$.  In the isotropic case, the spatial and temporal lattice spacing are identical, $a_\tau = a_s$. The requirement that the spatial extent should be larger than the temporal extent, $L\gg 1/T$ or $N_s\gg N_\tau$,  leads to the need for large lattices, with, say, $N_s \geq 4 N_\tau$.
In the anisotropic case, $a_\tau \ll a_s$, it is possible to use $N_s \sim N_\tau$, while keeping $L\gg 1/T$, due to the coarseness of the spatial lattice. Since simulations at different lattice spacings require a nontrivial tuning of the bare parameters, no continuum limit has been taken (yet). However, using the fixed-scale approach, it is easy to simulate at many different temperatures, simply by changing $N_\tau$ using the relation $T=1/(a_\tau N_\tau)$.

A handful of studies for $N_f=2+1(+1)$ flavours of Wilson quarks exist. The fixed-scale approach is used on isotropic lattices by the 
 Budapest-Wuppertal collaboration \cite{Borsanyi:2012uq,Borsanyi:2015waa}, for pion masses $m_\pi=545, 440, 285$ MeV, and taking a continuum extrapolation;
by  the WHOT collaboration  \cite{Umeda:2012er,Taniguchi:2016ofw,Kanaya:2019okb,Taniguchi:2020mgg}
using gradient flow at a single lattice spacing;
and using twisted mass fermions (with $N_f=2+1+1$ flavours)  \cite{Burger:2018fvb,Kotov:2020hzm,Kotov:2021rah}, for multiple pion masses, including at the physical point at a single lattice spacing. All of these studies use isotropic lattices.

\begin{figure}
\begin{center}
\bmp{8}
   \includegraphics[width=7cm]{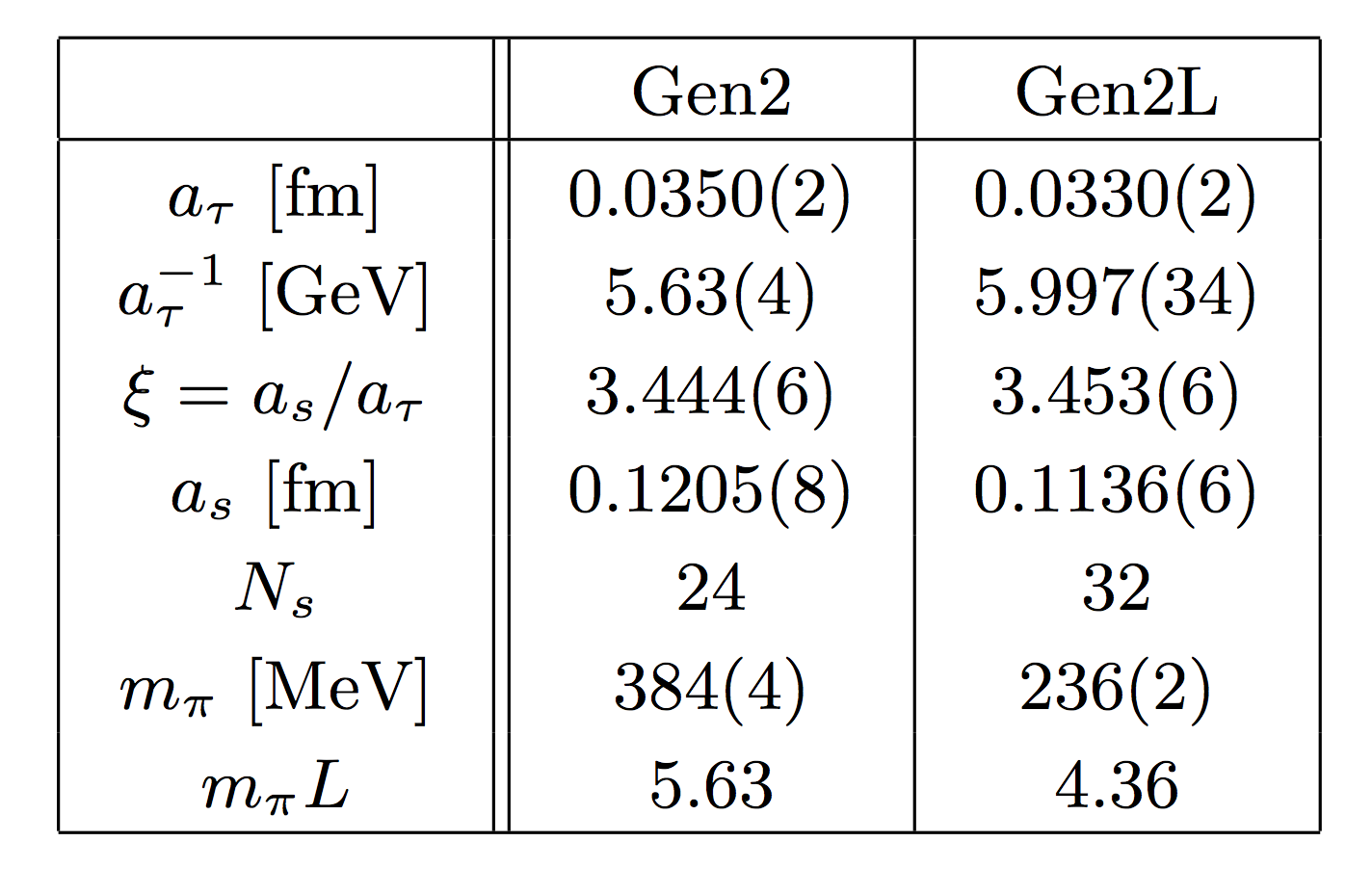}
   \includegraphics[width=7.5cm]{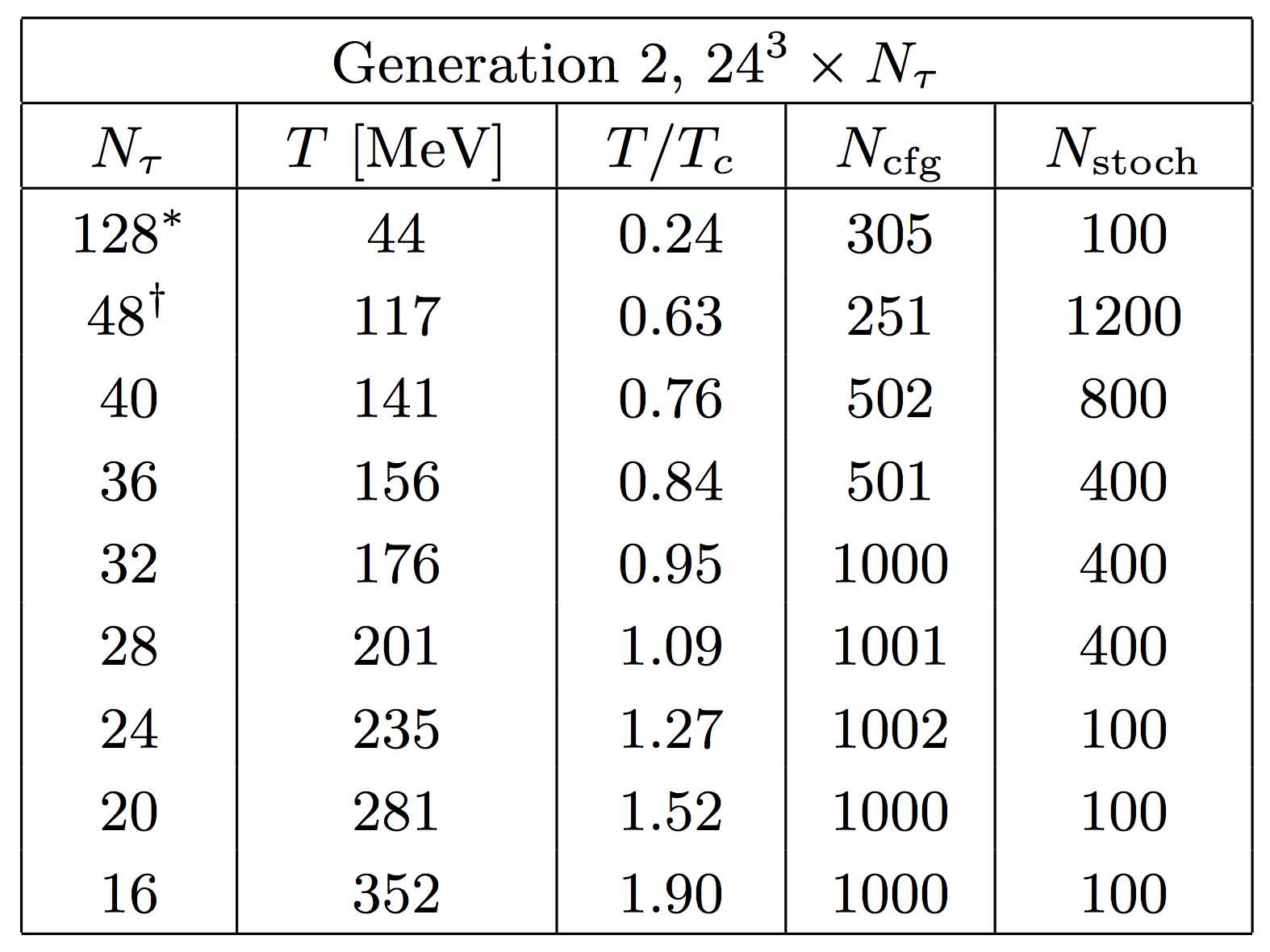} 
\emp
\bmp{7}
  \vspace*{1.65cm}
   \includegraphics[width=6.5cm]{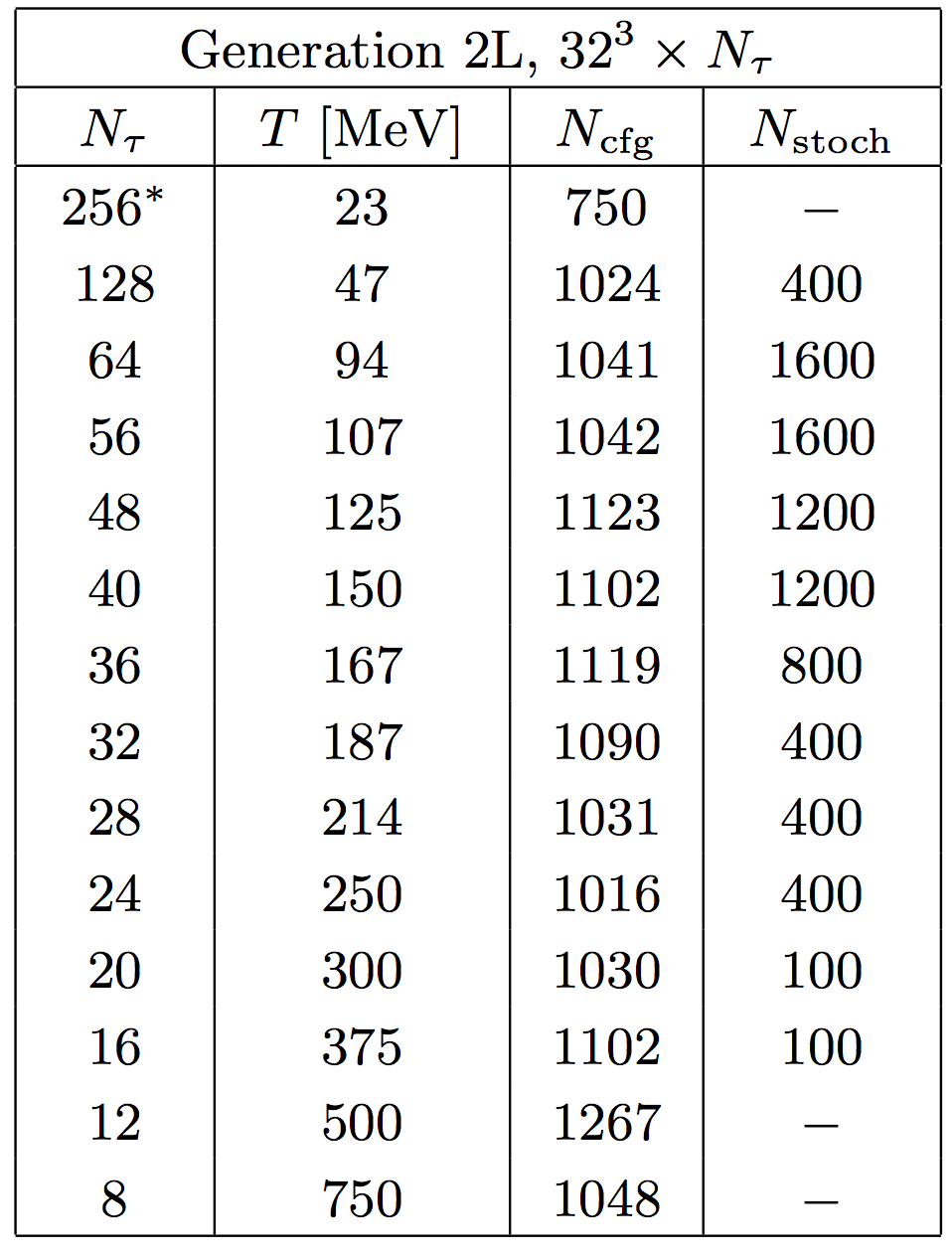}
\emp
\caption{Lattice details (top left) for Generation 2 and 2L, and details of the finite-temperature ensembles for Generation 2 (bottom left) and 2L (right) \cite{Aarts:2020vyb}.}
\label{fig:tables}
\end{center}
\end{figure}

In this contribution we concentrate mainly on results obtained by the FASTSUM collaboration, employing a fixed-scale approach on anisotropic lattices \cite{Aarts:2020vyb}. Simulations are carried out using a Symanzik-improved gauge action and a Wilson tadpole-improved clover fermion action, with stout-smeared links. Details of the parameters and ensembles are given in Fig.\ \ref{fig:tables}. The tuning of the bare parameters and the ensembles at the lowest temperature, indicated with a $^*$, have been constructed by the HadSpec collaboration \cite{Edwards:2008ja,HadronSpectrum:2008xlg,Cheung:2016bym,Wilson:2019wfr}. We note that the renormalised anisotropy, $\xi=a_s/a_\tau \approx 3.45$, and that the temporal cutoff $a_\tau^{-1} \approx 6$ GeV. Ensemble information is given for 2 generations: while the strange quark mass $m_s$ is at its physical value, the light quark mass  is larger than in nature. The main difference between the 2 generations is the pion mass, with $m_\pi=384(4)$ MeV (Generation 2) and $m_\pi=236(2)$ MeV (Generation 2L, L $=$ light). 
We emphasise that having many time slices available, also in the high-temperature phase, is useful for the analysis of temporal correlations functions, i.e.\ spectroscopy.
The (pseudo)critical temperature in Gen 2 has been determined using the renormalised Polyakov loop and is estimated to be $T^P_{\rm pc} = 185(4)$ MeV \cite{Aarts:2014nba}. The transition temperature in Gen 2L will be discussed in this contribution. One aim is to analyse the dependence on the pion mass, i.e.\ the change going from Gen 2 to Gen 2L. 

\section{Physical observables}

The dependence of the thermal transition on the light quark mass can be seen in observables involving the light degrees of freedom: (quark number) susceptibilities, the chiral condensate and the chiral susceptibility. The former are sensitive to the transition from hadronic to quark fluctuations, while the latter are relevant for chiral symmetry restoration.  In absence of a proper phase transition, details of the thermal transition depend on the observable and are hence characterised by observable-dependent pseudo-critical temperatures $T_{\rm pc}$. The expectation is that the $T_{\rm pc}$'s shift to lower values as $m_\pi$ is reduced; eventually  $T_{\rm pc}$'s  for different observables should coincide when the transition becomes a proper (chiral) phase transition.

\begin{figure}
\begin{center}
   \includegraphics[height=5.9cm]{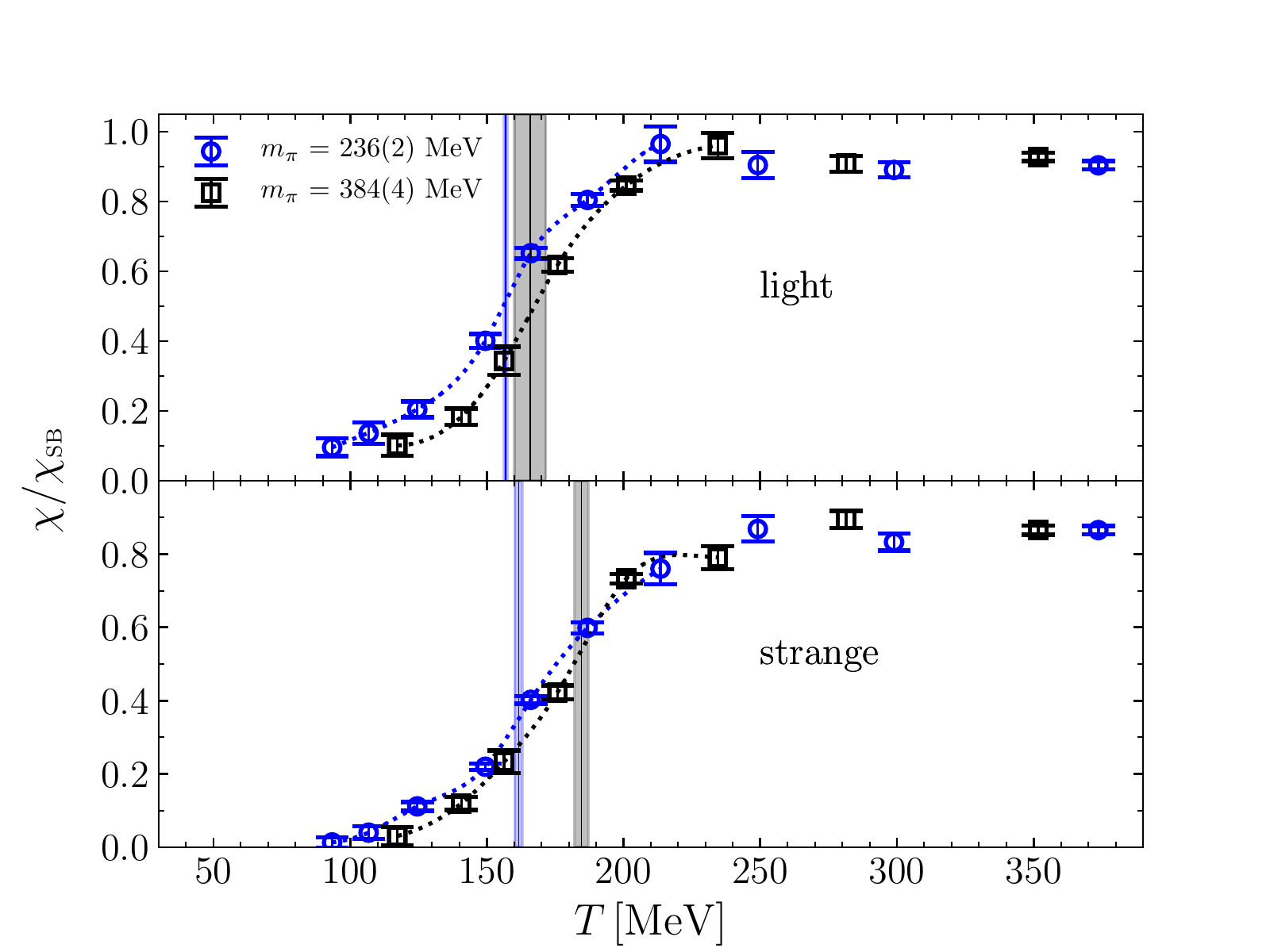} 
   \includegraphics[height=5.9cm]{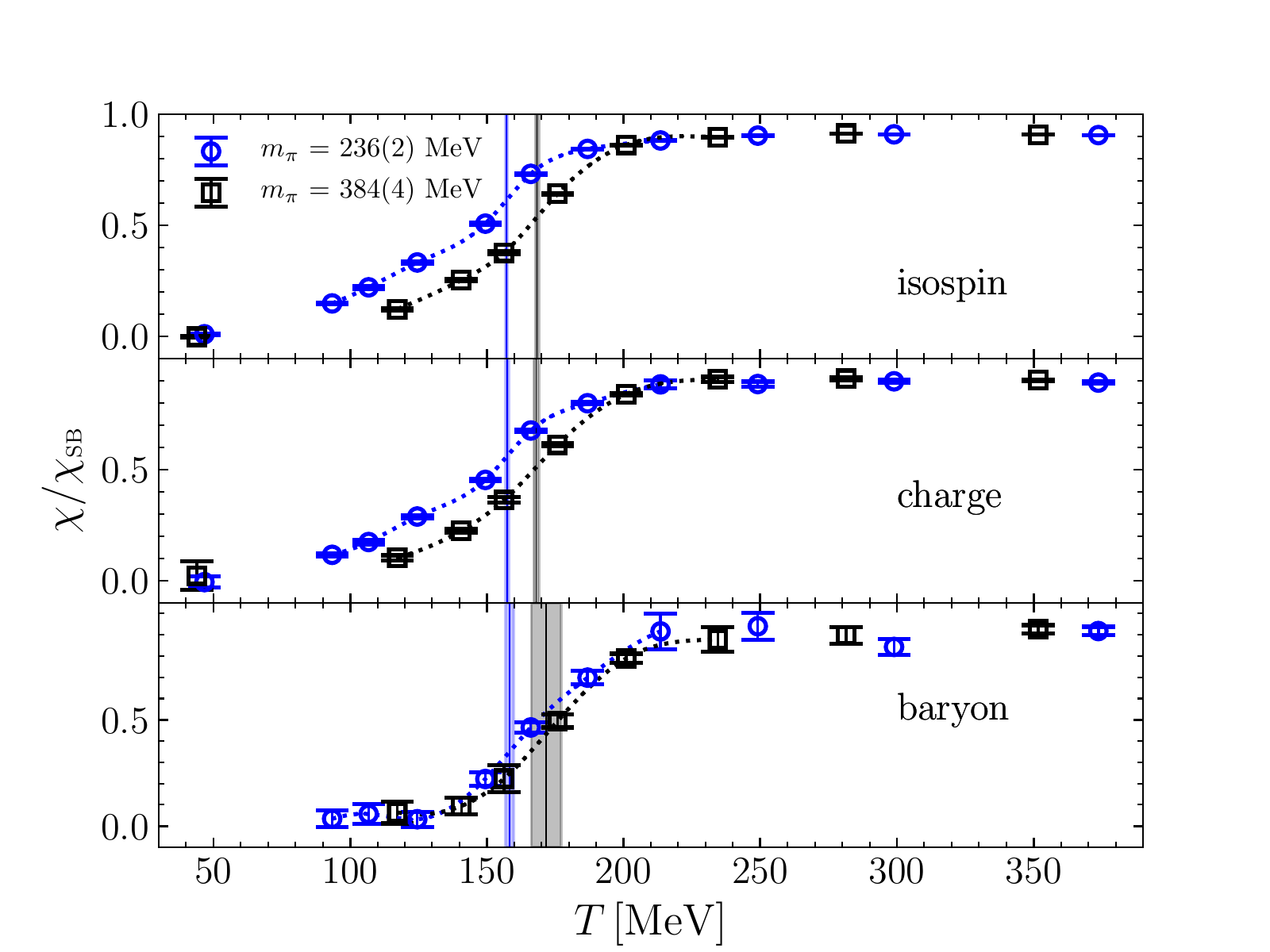}
\caption{Susceptibilities of light and strange quark number (left) and of isospin, electric charge and baryon number (right), normalised with the susceptibilities $\chi_{\rm SB}$ for massless fermions on a lattice in absence of interactions, as a function of temperature.
 The vertical bands indicate the inflection point for Gen 2 ($m_\pi=384(4)$ MeV) and Gen 2L ($m_\pi=236(2)$ MeV)  \cite{Aarts:2020vyb}.
}
\label{fig:sus}
\end{center}
\end{figure}

 In Fig.\ \ref{fig:sus} we show results for the fluctuations of light and strange quark number (left), and of isospin, electric charge and baryon number (right). These are normalised with the susceptibilities $\chi_{\rm SB}$ for massless fermions on a lattice in absence of interactions. 
Black symbols represent  Gen 2, i.e.\ the heavier pion, and blue symbols Gen 2L, i.e.\ the lighter pion. The transition is identified by the inflection point of cubic spline fits  and indicated with vertical bands; the thickness indicates statistical uncertainty. As the pion mass is reduced, a consistent shift of inflection points towards lower temperatures is observed. The actual values of $T_{\rm pc}$'s are collected in Table \ref{tab:tpc}, to be discussed later.

The chiral condensate contains information on chiral symmetry breaking/restoration. The bare condensate is subject to  additive and multiplicative renormalisation. Since we use a fixed-scale approach the renormalisation factors are  identical at all temperatures. Nevertheless we consider the renormalised chiral condensate. Following the Budapest-Wuppertal collaboration \cite{Borsanyi:2012uq}  (which follows Ref.\ \cite{Giusti:1998wy}), we consider 
\be
\label{eq:cc}
 m_R \bra\bar\psi\psi\ket_R(T) = \frac{\Delta_{\bar\psi\psi}^2(T) }{2 N_f \Delta_{PP}(T)} + \ldots
\ee
with the subtracted chiral condensate and  pseudoscalar susceptibility
\bea
&& \Delta_{\bar\psi\psi}(T) = \bra \bar \psi_l \psi_l \ket(T) - \bra \bar \psi_l \psi_l \ket(T=0), \\
&& \Delta_{PP}(T) =  \int d^4x\, \bra P(x) P(0)\ket(T) -  \int d^4x\, \bra P(x) P(0)\ket(T=0).
\eea
Here it is important to note that the LHS of Eq.\ (\ref{eq:cc}) is finite, while the RHS contains computable bare quantities. Finally we consider the 
dimensionless ratio,
\be
\frac{m_R \bra\bar\psi\psi\ket_R(T)}{m_\pi^2 m_\Omega^2},
\ee
which is finite in chiral limit.
Besides the condensate we also consider the subtracted susceptibility
\be
\Delta_{\chi_{\bar\psi\psi}} =\chi_{\bar\psi\psi}(T)-\chi_{\bar\psi\psi}(T=0).
\ee
It is noted that there is a remaining multiplicative renormalisation, which is however $T$ independent and hence only affects the vertical scale. 

\begin{figure}
\begin{center}
   \includegraphics[height=5.9cm]{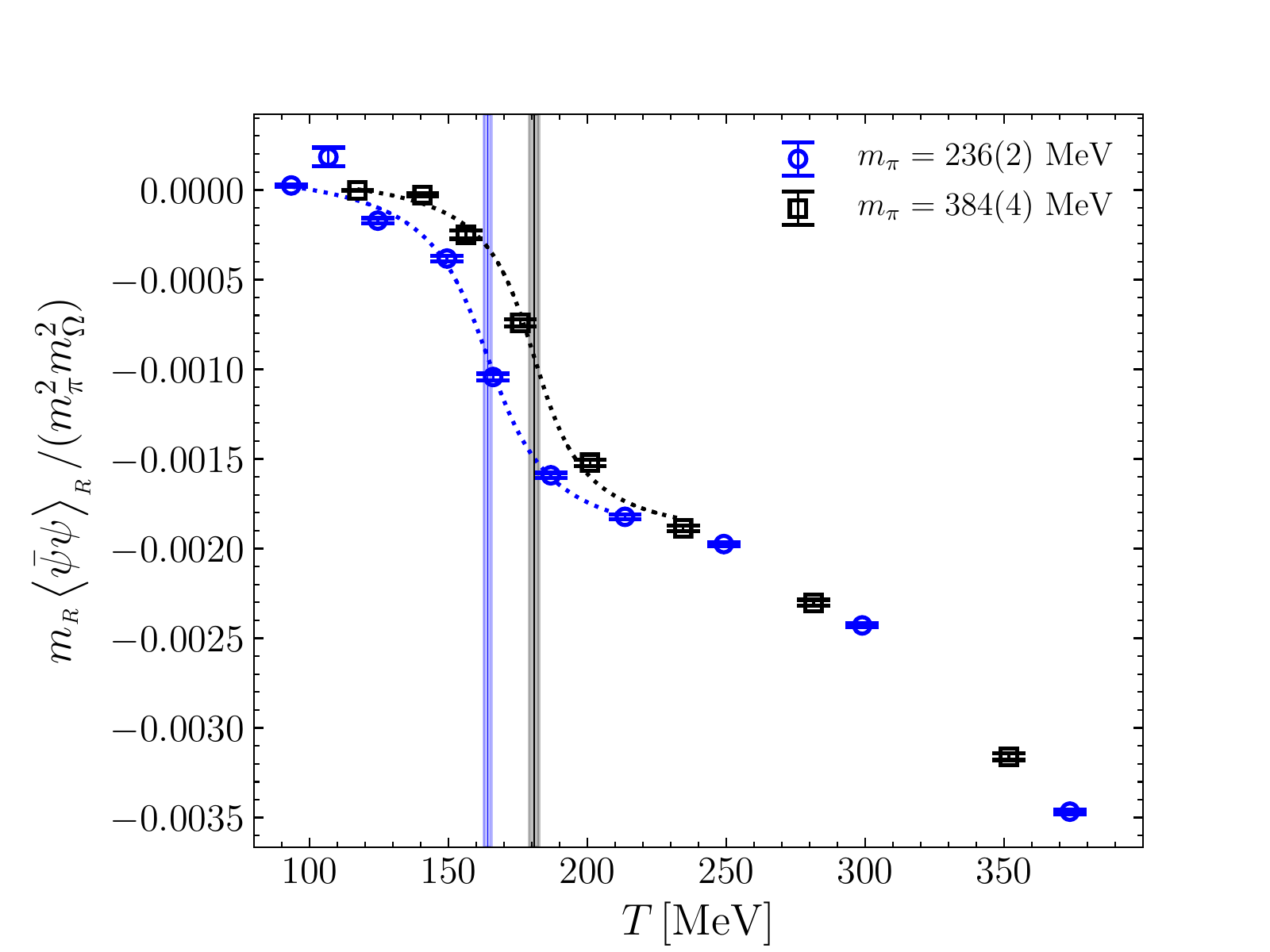}
     \includegraphics[height=5.9cm]{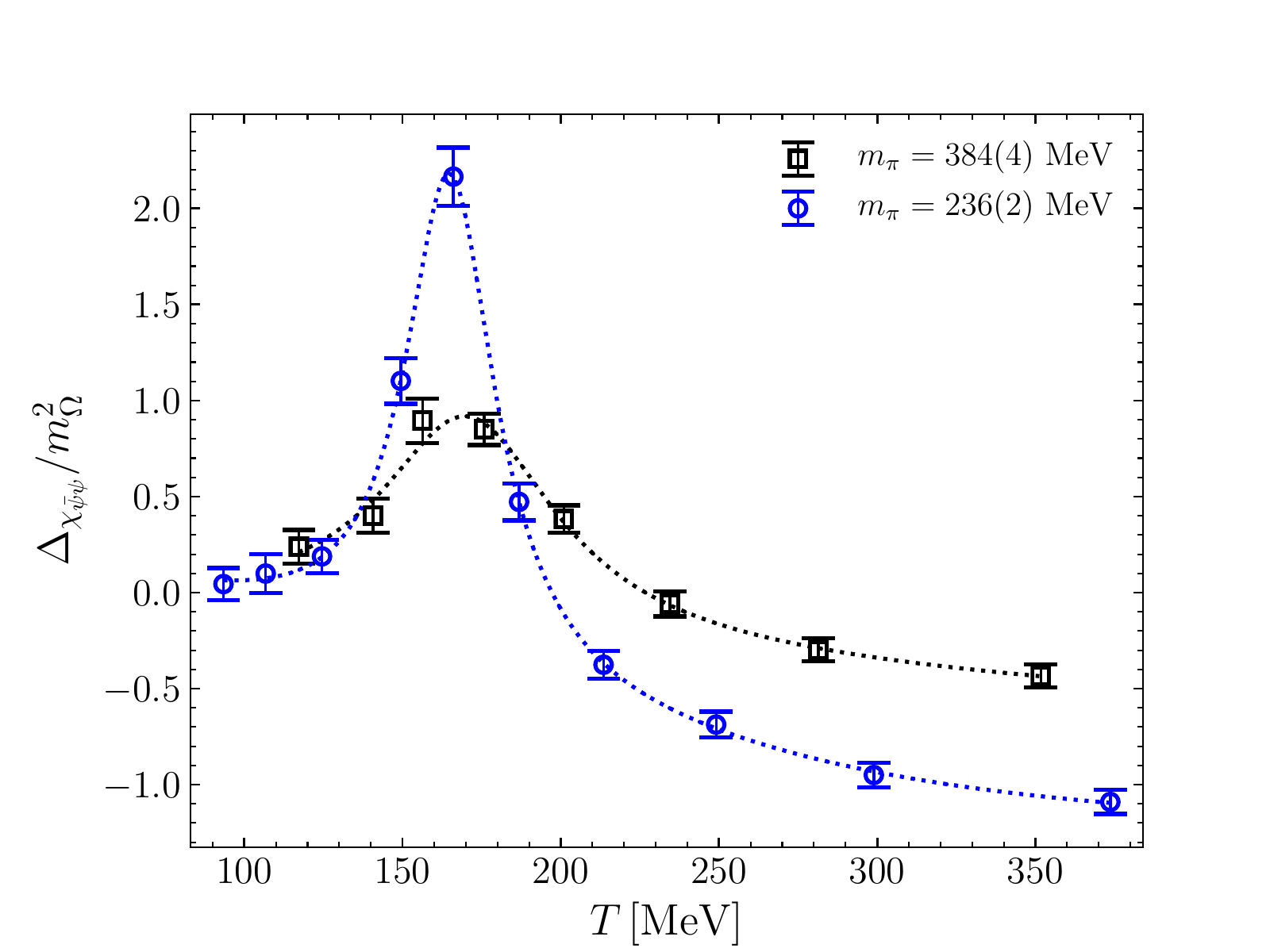}
\caption{Dimensionless renormalised chiral condensate (left) and subtracted chiral susceptibility (right) as a function of temperature.
The vertical bands on the left indicate the inflection points for Gen 2 and Gen 2L \cite{Aarts:2020vyb}.}
\label{fig:ch}
\end{center}
\end{figure}

The results for the chiral condensate and the susceptibility are shown in Fig.\  \ref{fig:ch}. 
For the chiral condensate we observe a shift of the transition region to lower temperature. 
For the chiral susceptibility, we note  in addition a  more pronounced peak for the lighter pion. For both observables estimates of $T_{\rm pc}$'s can be extracted and they are collected in Table \ref{tab:tpc}.

\section{Pseudo-critical temperatures and comparison with other approaches}

\begin{table}[t]
\begin{center}
    \begin{tabular}{| c|| c | c |  }
    \hline
                            & \multicolumn{2}{c|}{$T_{\rm pc}$ [MeV]}   \\
        \hline
            & Gen2L & Gen2 \\        \hline   
         observable    & $m_\pi = 236(2)$ MeV& $m_\pi = 384(4)$ MeV \\        \hline   
        $\chi_{\rm light}$   	 & $157(1)$      & $166(6)$ \\
        $\chi_{\rm strange}$     & $162(2)$      & $184(3)$ \\
        $\chi_{\rm I}$       	 & $157.2(4)$  & $168.4(6)$ \\
        $\chi_{\rm Q}$       	 & $157.5(6)$  & $168.1(6)$ \\
        $\chi_{\rm B}$       	 & $158(2)$      & $172(5)$ \\
        $\bra\bar\psi\psi\ket_R$ & $164(2)$      & $181(2)$ \\
        $\chi_{\bar\psi\psi}$    & $165(2)(2)$   & $170(3)(2)$ \\
            \hline
    \end{tabular}
    \caption{Estimates of the pseudocritical temperatures from a range of fermionic observables, for both values of the pion mass.}
    \label{tab:tpc}
\end{center}
    \end{table}

\begin{figure}
\begin{center}
   \includegraphics[height=5.9cm]{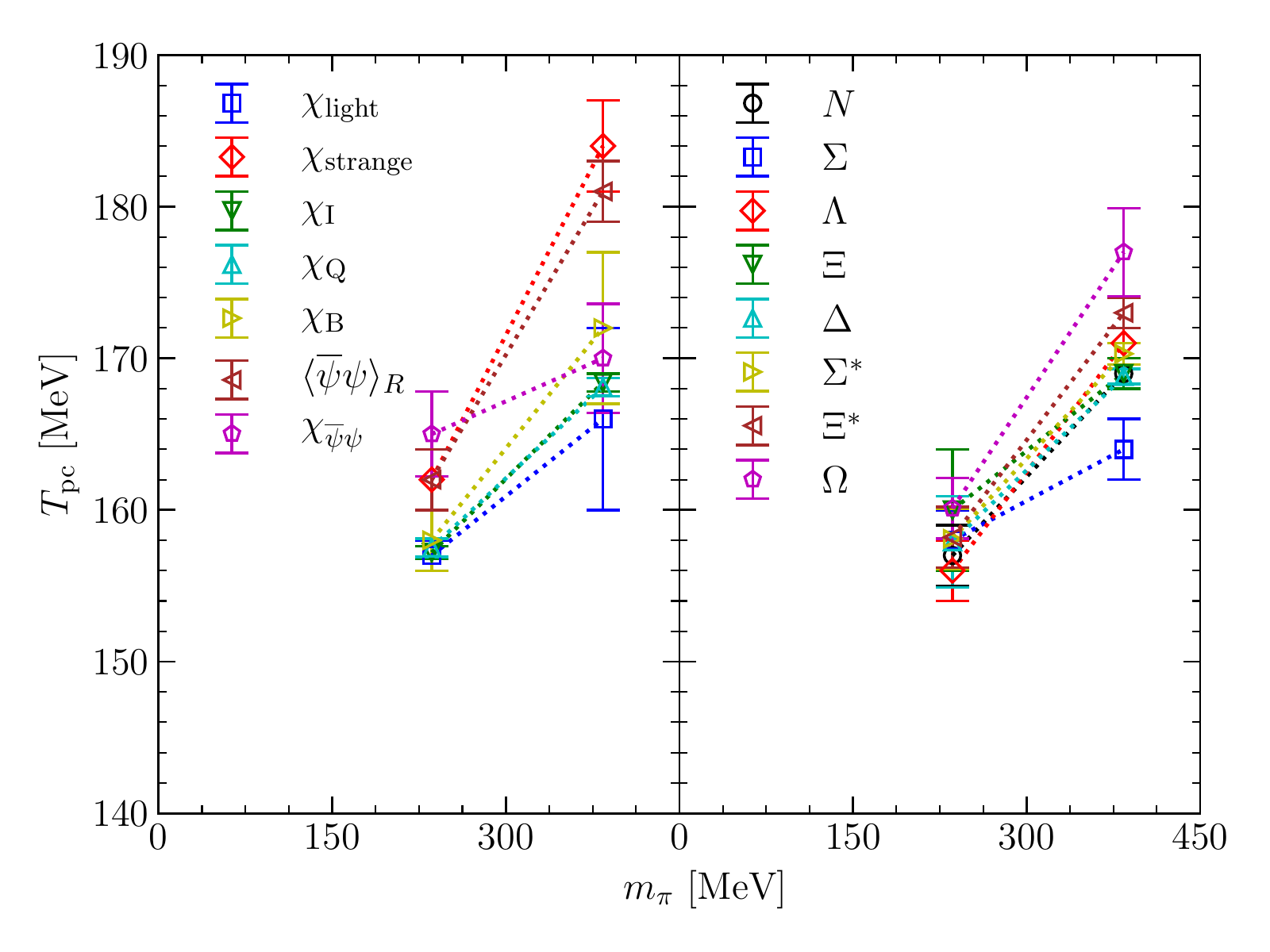}
     \includegraphics[height=5.9cm]{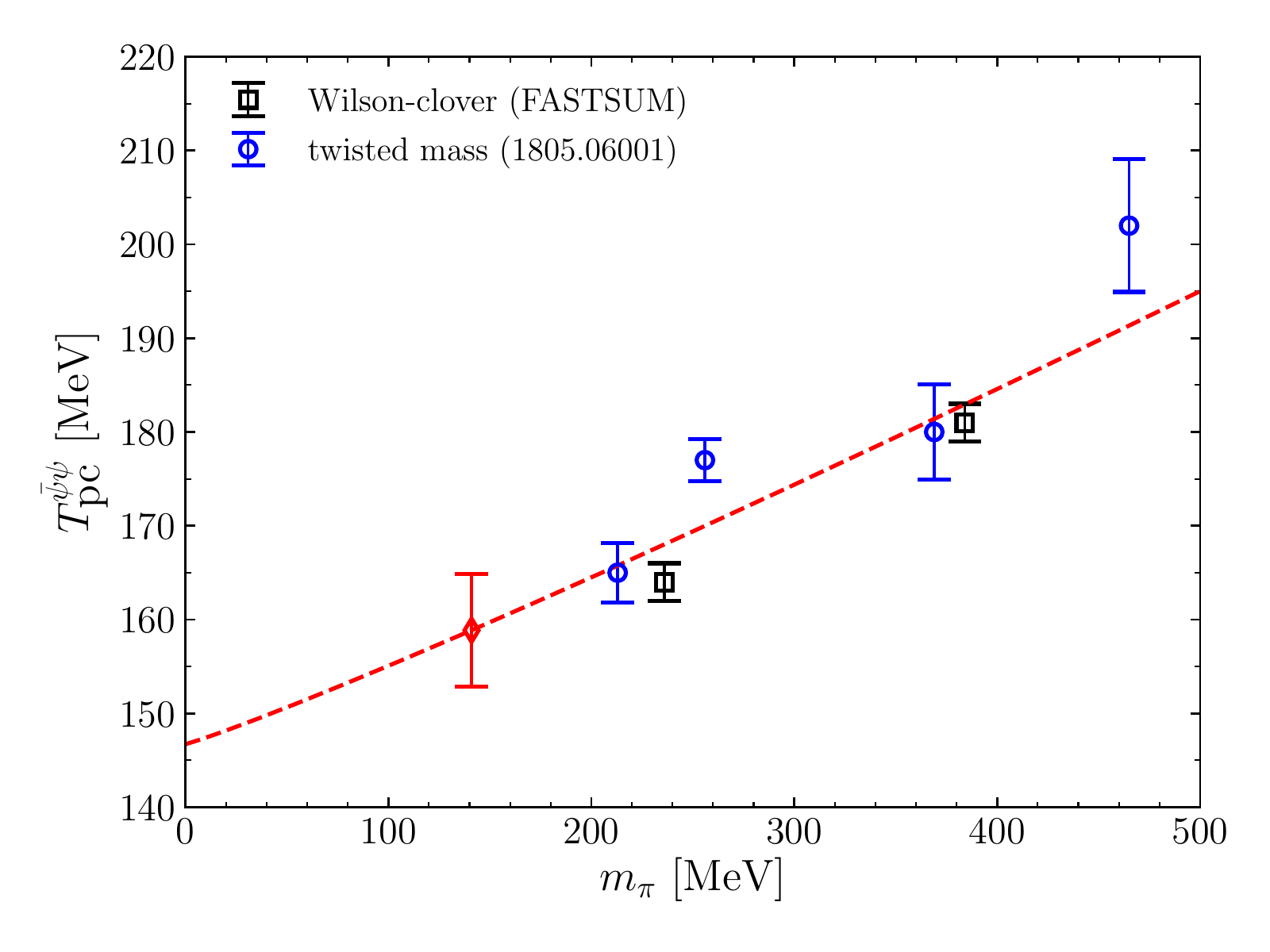}
\caption{Left: Pseudocritical temperatures as a function of the pion mass, as obtained from susceptibilities and the chiral condensate (left pane) and from baryon parity doubling (right pane).
Right: Pseudocritical temperatures as obtained from the inflection point of the renormalised chiral condensate, determined by twisted-mass fermions and this work. The fit given by the dashed line is discussed in the text;  the diamond indicates the extrapolated value at the physical point \cite{Aarts:2020vyb}. }
\label{fig:tpc}
\end{center}
\end{figure}

The estimates of the pseudocritical temperatures for the observables considered so far are collected in Table \ref{tab:tpc}, for both values of the pion mass. They are also shown in Fig.\ \ref{fig:tpc} (most left pane). We observe that for the lighter pion the spread in temperature is reduced. This may be a sign of being closer to a proper phase transition. We also note that $\chi_{\bar\psi\psi}$ is somewhat of an outlier at the heavier pion, which may be due to the peak being quite broad in that case.

One may wonder whether anything can be said about the physical point or even massless quarks. An extrapolation using only two pion masses is somewhat audacious, but it is possible to compare with other fermion formulations. Here we consider in particular twisted-mass fermions, with $N_f=2+1+1$ \cite{Burger:2018fvb}, with a similar range of pion masses. The estimates for the pseudocritical temperature as obtained  from the inflection point of the renormalised chiral condensate are shown in Fig.\ \ref{fig:tpc} (right). 
We may perform a combined extrapolation to physical point, using the Ansatz
\be
T_{\rm pc}^{\bar\psi\psi}(m_\pi) = T_0 + \kappa m_{\pi}^{2/\Delta} \qquad\qquad
\Delta = 1.833  \; \mbox{fixed},
\ee
(this value of $\Delta$ corresponds to the three-dimensional O(4) prediction).
We find that $T_0 =  147(4)$ MeV and hence obtain at the physical point
\be
T_{\rm pc}^{\bar\psi\psi} = 159(6) \; \mbox{MeV} \qquad\qquad \mbox{(physical point)}.
\ee
This result is indicated with the diamond in Fig.\ \ref{fig:tpc} (right). We may compare this with the results obtained using  simulations of 
staggered quarks directly at the physical point, 
$T_{\rm pc}^{\bar\psi\psi} =155(3)(3)$ MeV -- Budapest-Wuppertal \cite{Borsanyi:2010bp} -- and 
$T_{\rm pc} = 156.5(1.5)$ MeV  -- hotQCD  \cite{HotQCD:2018pds} -- although we hasten to state that in our case no continuum extrapolation has been carried out.

\section{Baryons and parity doubling}
 
Chiral symmetry breaking plays an important role in the hadronic spectrum at $T=0$, for light mesons but also for baryons. In particular, the operators representing baryons come in pairs, related by parity. In nature, however, the corresponding states are not degenerate and there is no particle doubling in the QCD spectrum. As an example, consider the nucleon groundstate: 
the positive-parity state has mass $m_+ = m_N = 0.939$ GeV, while the negative-parity state is heavier, with $m_-=m_{N^*} = 1.535$ GeV. The absence of parity doubling is a direct consequence of chiral symmetry breaking.

\begin{figure}[t]
\begin{center}
   \includegraphics[height=5.9cm]{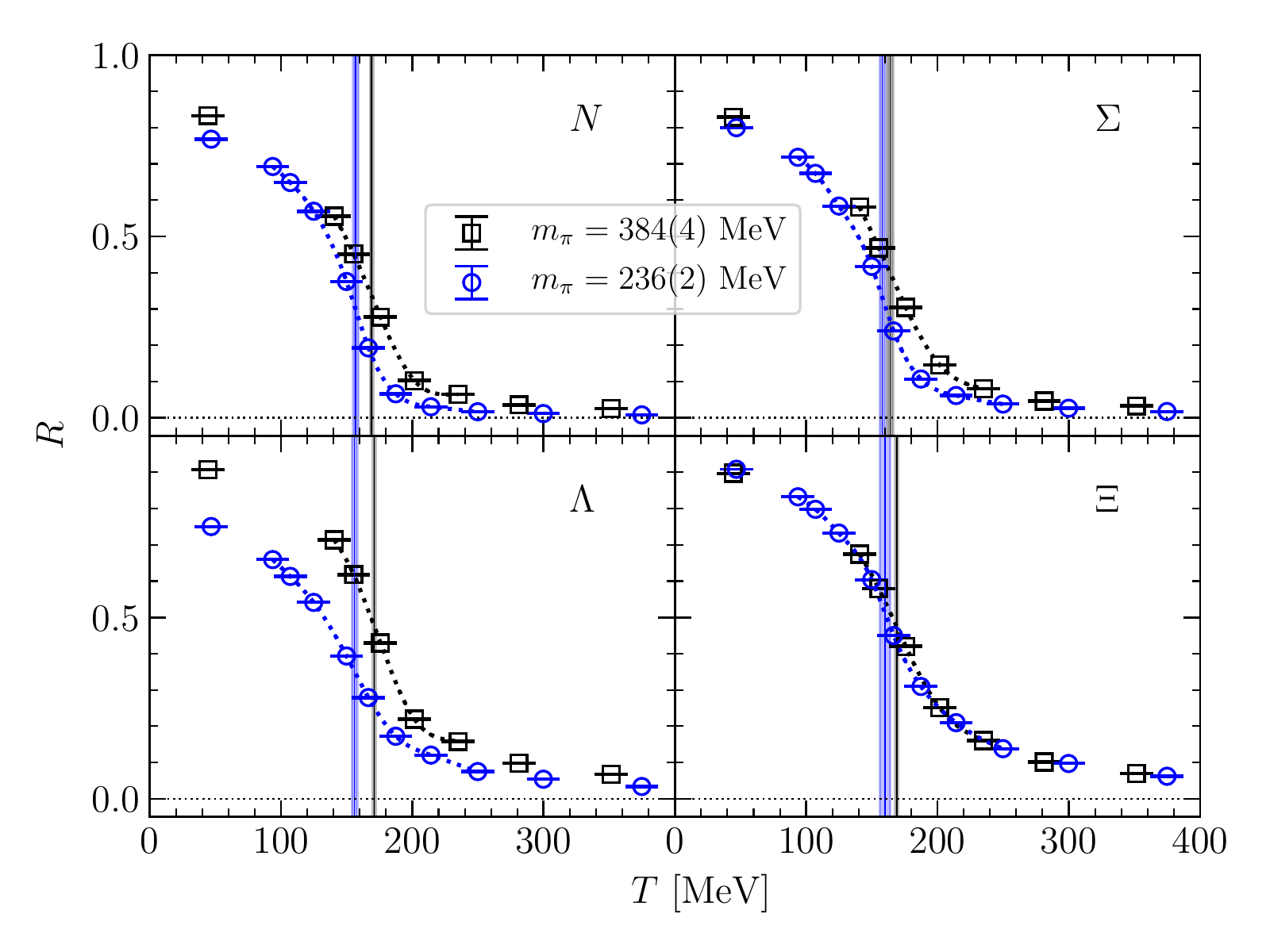}
   \includegraphics[height=5.9cm]{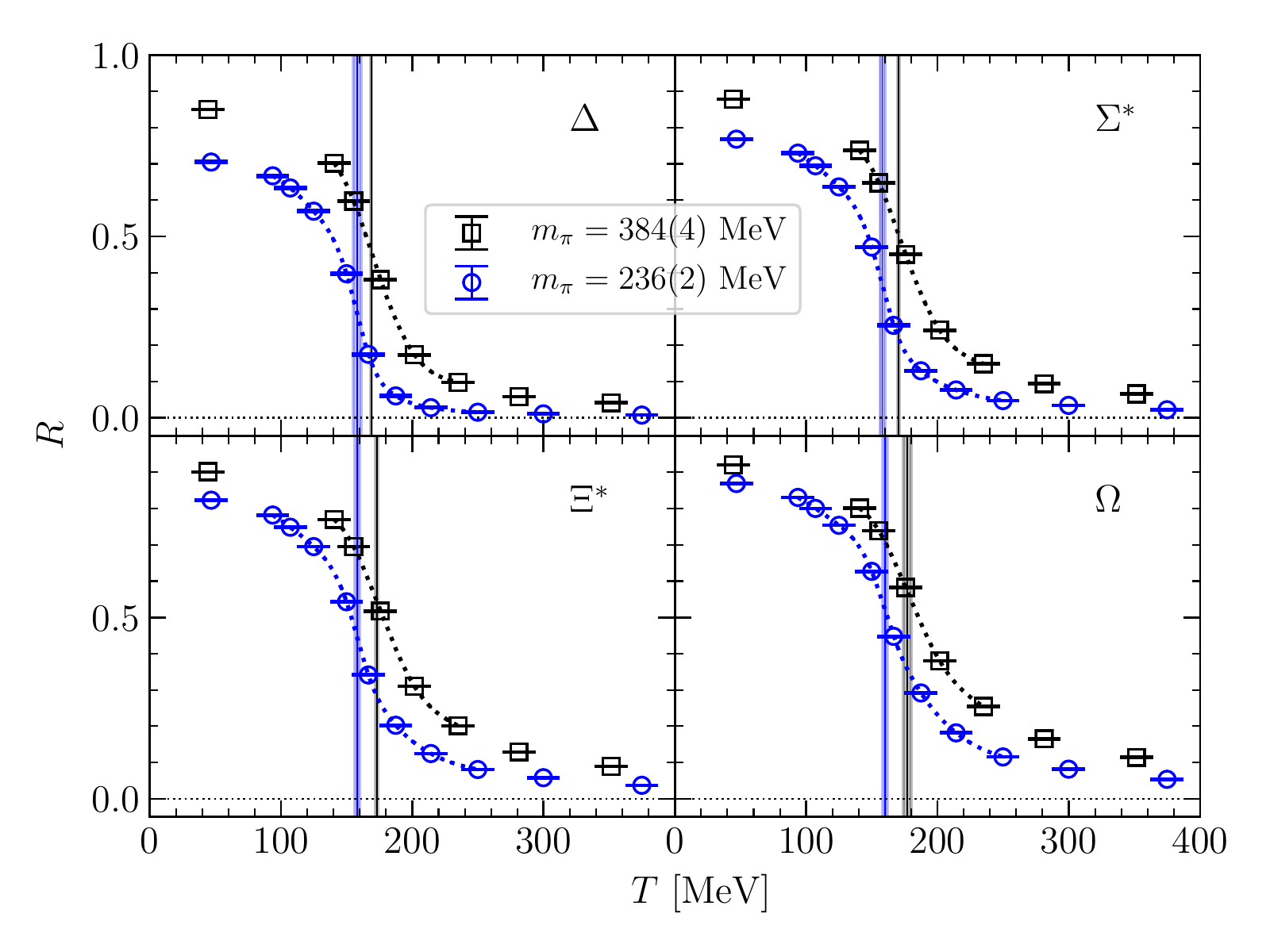}
\caption{Parity-doubling $R$ parameter as a function of temperature for octet (left) and decuplet (right) baryons.
The vertical bands indicate the inflection points for Gen 2 and Gen 2L \cite{Aarts:2020vyb}.
 }
 \label{fig:R}
\end{center}
\end{figure}

One may wonder what happens at nonzero temperature when chiral symmetry is restored eventually. If the restoration takes place in the quark-gluon plasma phase, most likely the light mesons and baryons have disappeared from the spectrum. Nevertheless, a question to be asked is whether  there is a precursor to chiral symmetry restoration in the hadronic phase and hence an onset of parity doubling as well. 
To study this, we have investigated in-medium effects for the positive- and negative-parity baryonic groundstates in a series of papers 
   \cite{Aarts:2015mma,Aarts:2017rrl,Aarts:2018glk}. We observed that there is an emerging degeneracy around the crossover temperature, with the negative-parity states becoming lighter as the temperature increases while the positive-parity states are seen to be nearly temperature independent.
 
Since it is hard to define a groundstate at nonzero temperature, and especially in the high-temperature phase, it is important to realise that 
 parity doubling can already be seen at the level of baryonic correlation functions. Following Ref.\ \cite{Datta:2012fz}, we may consider the ratio
 \be 
R(\tau) = \frac{ G_+(\tau)-G_-(\tau)}{ G_+(\tau) + G_-(\tau)},
\ee
where $G_\pm(\tau)$ is the positive/negative-parity baryon correlator, and construct a quasi-order parameter
\be
R = \frac{\sum_n R(\tau_n)/\sigma^2(\tau_n)}{\sum_n 1/\sigma^2(\tau_n)},
\ee
summed over the time slices (and weighted with the estimated error).
In absence of parity doubling and when the positive-partner groundstate is much lighter than the negative-partner one, one finds that $R\sim 1$, while in presence of parity doubling $R=0$. Details can be found in Refs.\ \cite{Aarts:2015mma,Aarts:2017rrl,Aarts:2018glk}.

\begin{table}[t]
\begin{center}
\begin{tabular}{|c|| cccc| }
\hline
 \; $T_{\rm inf}$[MeV] \;	 & $N$ & $\Sigma$ & $\Lambda$ & $\Xi$  	\\
 \hline
 Gen2 & 169(1) & 164(2) & 171(1) & 169(1)  	\\
 Gen2L &  157(2) & 158(2) & 156(2) & 160(4)   \\
\hline
 $T_{\rm inf}$[MeV]	 &  $\Delta$ & $\Sigma^*$ & $\Xi^*$	& $\Omega$ 	\\
 \hline
 Gen2 & \; 168.8(5) \; & \; 170.3(7) \; & \; 173(1)	 \;& \; 177(3) \;	 	\\
 Gen2L &   158(3) & 158(2) & 158(2) & 160(2) \\
\hline
\end{tabular}
\caption{Estimates of the transition temperatures obtained from the $R$ parameter for the baryon channels.}
\label{tab:T_R}
\end{center}
\end{table}

 In Fig.\ \ref{fig:R} we show the parity-doubling $R$ parameter as a function of temperature for octet and decuplet baryons. It is noted that $R$ indeed changes from being ${\cal O}(1)$ at low temperature to being close to zero at high temperature. The presence of the strange quark, with a nonnegligible mass,  is visible in the high-temperature phase, especially in the $\Xi^{(*)}$ and $\Omega$ channels, with strangeness 2 and 3.  The effect of reducing the light quark masses, going from Gen 2 to Gen 2L, is equally visible for the $N$ and $\Delta$ channels  in the high-temperature phase. 
 
As before we observe a shift of the transition region to lower temperature, indicated by the vertical bands representing the inflection points for Gen 2 and Gen 2L. The actual values for the temperature of the inflection points are given in Table \ref{tab:T_R} and presented in Fig.\ \ref{fig:tpc} (left figure, right pane). 
As before, we note a reduced spread of points, again possibly indicating the presence of a real phase transition at even smaller pion masses.

\section{Summary}

In the past 15 years or so a detailed understanding of the thermal transition in QCD has been obtained using numerical simulations, focussing in particular  on thermodynamic observables. State-of-the-art studies use staggered quarks, directly at the physical point, and take the continuum limit. 
In this contribution the attention was instead on Wilson quarks, for which simulations are not yet simultaneously at the physical point and in the continuum limit. Nevertheless, a consistent picture is seen to emerge. 

Chiral symmetry restoration has implications for the hadronic spectrum at nonzero temperature. Wilson fermions on anisotropic lattices are particularly well suited to study this. To demonstrate this, we showed results for parity doubling of light and strange baryons, providing a complimentary window into the thermal transition region.
We stress that all results shown here have been obtained at a single lattice spacing, which indicates an important step for the future.

\ack

Results presented in this work were obtained with support from STFC via grants ST/L000369/1 and ST/P00\-055X/1, the Swansea Academy for Advanced Computing, SNF, ICHEC, the European Research Council (ERC) under the European Union's Horizon 2020 research and innovation programme under grant agreement No 813942. SK is supported by the National Research Foundation of Korea under grant NRF-2018R1A2A2A05018231 funded by the Korean government (MEST). 
We are grateful to DiRAC, HPC Wales, PRACE and Supercomputing Wales for the use of their computing resources. Results presented in this work were obtained using the PRACE Marconi-KNL resources hosted by CINECA, Italy and the DiRAC Extreme Scaling service and Blue Gene Q Shared Petaflop system at the University of Edinburgh operated by the Edinburgh Parallel Computing Centre. The DiRAC equipment is part of the UK's National e-Infrastructure and was funded by UK's BIS National e-infrastructure capital grant ST/K000411/1, STFC capital grants ST/H008845/1 and ST/R00238X/1, and STFC DiRAC Operations grants ST/K005804/1, ST/K005790/1 and ST/R001006/1.

\end{document}